\documentclass[]{aa}  

\usepackage{graphicx,natbib,journal_names}
\usepackage[breaklinks,colorlinks,urlcolor=blue,citecolor=blue,linkcolor=blue]{hyperref}
\usepackage{multirow,afterpage,pdflscape,lipsum,capt-of,mathtools}
\pdfoutput=1

\usepackage{txfonts}
\usepackage{soul}
\usepackage{array}
\usepackage{rotating}
\newcommand{\RHK}{R$^\prime_\mathrm{HK}$}
\newcommand{\lRx}{$\log \mathrm{R}_\mathrm{X}$}

\newcommand{\lRHK}{$\log \mathrm{R}^\prime_\mathrm{HK}$}

\newcommand{\alfB}{$\alpha_\mathrm{mm}$}

\newcommand{\TB}{$\mathrm{T_B}$}
\newcommand{\teff}{T$_\mathrm{eff}$}
\newcommand{\Tbp}{T$_\mathrm{B}(\nu)$}
\usepackage{colortbl}
\definecolor{Grey}{rgb}{0.5,0.5,0.5} 
\definecolor{MyRed}{rgb}{0.9,0.0,0.0} 
\definecolor{MyPink}{rgb}{0.8,0.3,0.5} 
\definecolor{MyMediumBlue}{rgb}{0.7,0.72,1.0} 
\definecolor{MyGreen}{rgb}{0.0,0.9,0.5} 
\definecolor{SWGreen}{rgb}{0.0,0.5,0.0} 
\definecolor{SWRed}{rgb}{0.9,0.0,0.0} 
\definecolor{AMpurple}{rgb}{0.9,0,1.0} 
\definecolor{AMbrown}{rgb}{0.8,0.3,0.3}

\begin{document}

    \title{EMISSA - Exploring Millimetre Indicators of Solar-Stellar Activity} 
   \subtitle{II. Towards a robust indicator of stellar activity}
    \titlerunning{EMISSA II. Towards a robust stellar activity indicator}
    
   \author{A. Mohan\inst{1,2}
         , S. Wedemeyer\inst{1,2}
         , P. H. Hauschildt\inst{3}
         , S. Pandit\inst{1,2}
         \and
         M. Saberi\inst{1,2}
          }
        \institute{Rosseland Centre for Solar Physics, University of Oslo, Postboks 1029 Blindern, N-0315 Oslo, Norway
    \and
    Institute of Theoretical Astrophysics, University of Oslo, Postboks 1029 Blindern, N-0315 Oslo, Norway\\
    \email{atulm@astro.uio.no}
    \and
    Hamburger Sternwarte, Gojenbergsweg 112, 21029 Hamburg, Germany
    }

             
             
\authorrunning{Mohan, A. et al.}

   \date{Received 30 June, 2022 / Accepted  7 August, 2022  }

 
  \abstract
  {
  {An activity indicator, which can provide a robust quantitative mapping between the stellar activity and the physical properties of its atmosphere, is important in exploring the evolution of the observed active phenomena across main-sequence stars of different spectral types.
  Common activity indicators do provide qualitative correlations with physical properties such as \teff\ and the rotation period, among others. However, due to the large variability in their values, even for a single star, defining robust quantitative mappings between activity and physical properties is difficult.
  Millimetre (mm) wavelengths probe the different atmospheric layers within the stellar chromosphere, providing a tomographic view of the atmospheric dynamics. 
  } }
 {The project aims to define a robust activity indicator by characterising mm brightness temperature spectra (\Tbp) of the cool main-sequence stars (\teff\ $\sim$ 5000 -- 7000\,K) compiled by paper I in this series. The sample contains 13 stars, including the Sun.}
    {We derived the mm \Tbp\ spectral indices (\alfB) for cool stars, including the Sun, based on observations in the 30 -- 1000\,GHz range. The derived values for \alfB\ are explored as a function of various physical parameters and empirical power-law functions were derived. { We also compared \alfB\ estimates with other activity indicators.}}
  {
  Despite the estimation errors, \alfB\ values could distinguish the cool stars  well, unlike common activity indicators.  
  The low estimation errors on the derived trends of \alfB\ versus physical parameters suggest that \alfB\ could be a robust activity indicator.
}
   {We note that \alfB, which is linked to chromospheric thermal stratification and activity in cool stars, can well distinguish and physically characterise the stars more robustly than common activity indicators. 
   {We emphasise the need for multi-frequency data across the mm band for stars, 
with a range of physical parameters and gathered at multiple epochs during { their} activity cycles. This will help to explore \alfB\ in a statistically robust manner and to study the emergence of chromospheric heating on the main sequence.}}


   \maketitle
%

\section{Introduction}
\label{sec:intro}
{In recent years, there has been a growing interest in the atmospheric activity of cool main-sequence stars (cool stars; effective temperature, \teff\,$\sim$3000 - 7000\,K) especially due to the links between activity, exo-space weather, and exo-planet habitability \citep[e.g.][]{2005AsBio...5..587G, Vidotto13_highB_forclosebyplanets,2019ApJ...877..105M,2020MNRAS.494.3766O}.
Cool stars of spectral types F -- M together host the largest number of Earth-like exoplanets in  habitable zones \citep{Bashi20_occurrence_of_smallplanetsFGK}. 
However, owing to their outer convection layer, these stars have active atmospheres that drive flares, eruptions, and other high-energy phenomena, which can potentially 
affect   
the atmospheres of nearby planets in habitable zones \citep{2007AsBio...7..185L,2010Icar..210..539Z}.
Ground- and space-based instruments provide plenty of data on stellar atmospheric emission primarily in X-ray to optical wavebands,  owing to large dedicated surveys \citep[e.g.][]{1997A&A...323L..49P,HARPS2014_datarelease,GaiaDR32021}. 
Based on the X-ray to optical observables, which are sensitive to different stellar atmospheric phenomena and layers, different activity indicators have been constructed. 
Commonly used indicators include the ratio of the Ca\,II H-K flux to bolometric flux (\RHK, \citet{Noyes84_RHK}) and the X-ray to bolometric flux ratio ($\mathrm{R_X}$). 
These activity indicators show trends with several physical parameters such as \teff, the rotation period, magnetic field strength, metallicity, among others, sometimes with complex multi-dimensional dependencies \citep[e.g.][]{Noyes84_RHK,stepien94_Defn_Rx+Ro_Vs_activity_n_manyCorCurves,2014MNRAS.444.3517M, Vidotto14_B_Vs_age_n_rot}. 
Though these trends provide a qualitative physical picture of the physical dependencies of activity, the large value ranges of these indicators make it hard to derive robust quantitative relations.   
For instance, \citet{Pace13_RHK_Drastic_variability} showed an age-dependent bimodality  of 
\RHK\ values for cool stars of same \teff. Even the values for a single star varied by factors of $2 - 4$, making it hard to distinguish the cool stars (\teff\ $\sim$ 4000 - 7000\,K) along the \RHK\ axis. 
Similar studies on $\mathrm{R_X}$ values have revealed variations as large as an order of magnitude in Sun-like stars~\citep[e.g.][]{2003A&A...404..637M}.
These studies show that the common indicators cannot provide robust quantitative scales which are essential to better constrain the atmospheric physical models and gain physical insights on the emergence of different levels of activity in stars of different 
types.}

\begin{figure*}[!t]
\centering
\includegraphics[width=\textwidth, height=0.28\textheight]{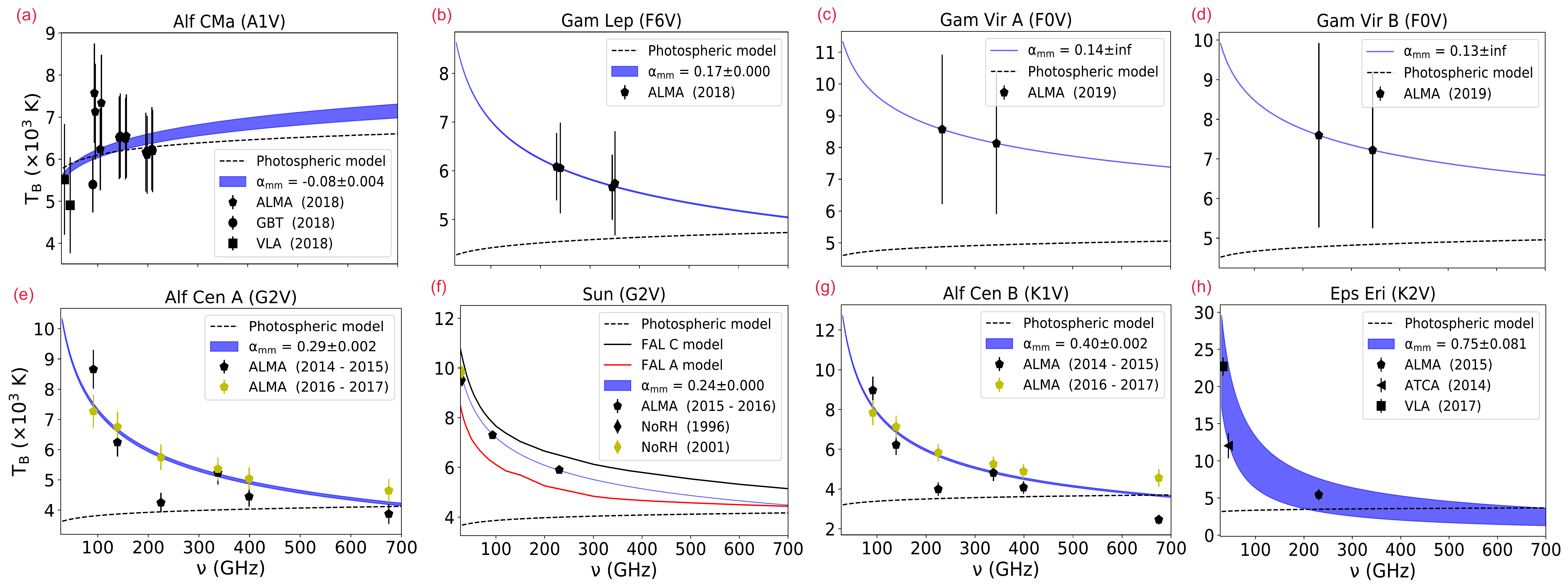}
\vspace*{-6mm}
\caption{mm-\Tbp\ in the 30 -- 1000\,GHz range. Titles show the names and spectral types. Family of power-law functions consistent with the \alfB\ estimation errors (shaded regions) and photospheric models (dotted line) are shown. For the Sun, chromospheric FAL models are shown. Multi-epoch data at $\nu$ are coloured differently.}
\label{fig1:fits}
\vspace{-0.4cm}
\end{figure*}

Observations and related models suggest that the atmospheric structure of cool stars undergoes significant changes as function of spectral type or equivalently in the \teff\ range 3000 -- 7000\,K \citep[e.g.][]{Donati09_Rev_Bfield,Linsky16_Stellar_chromRev}. 
{Being the middle layer of the stellar atmosphere, chromosphere plays a crucial role in the generation and propagation of active phenomena across the stellar atmosphere} \citep[for a review, see][]{Linsky16_Stellar_chromRev}.
It is hence highly desirable to construct a new and more reliable observational indicator of the chromospheric structure of cool stars, which is closely linked to the observed activity.
Continuum emission in the millimetre (mm) range ($\sim$30 -- 1000\,GHz) originates at different heights in the stellar chromosphere primarily due to thermal bremsstrahlung, which is formed under local thermal equilibrium (LTE) conditions \citep{Sven16_ALMA_science}.
Owing to its high sensitivity, ALMA enables the detection of chromospheric mm emission for a larger number of potential stellar targets than possible before. 
Several recent solar and stellar mm observations have demonstrated the unique tomographic potential of the mm-brightness temperature spectrum (mm-\Tbp) to facilitate the deduction of the chromospheric thermal structure and dynamics at various atmospheric 
layers 
in cool stars \citep[e.g.][]{2017SoPh..292...88W,White20_MESAS,2018MNRAS.481..217T}.
In the previous paper in this series \citep[Paper I;][]{2021A&A...655A.113M}, the mm-\Tbp\ spectrum was derived for main-sequence stars using archival ALMA data. 
The sample was made by systematically avoiding stars with potential flux contamination from known debris disks or unresolved companion stars.
Assuming that the observed  stellar mm flux can be approximated with a LTE Rayleigh-Jeans spectrum, \Tbp\ is expected to reflect the atmospheric thermal stratification.
The mm-\Tbp\ of every cool star in the sample significantly deviated from a corresponding purely photospheric model \citep[PHOENIX;][]{1999JCoAM.109...41H} spectrum, with the deviation from the photospheric model increasing as $\nu$ decreased (Fig.~\ref{fig1:fits}). The rise of $T_B$ with decreasing $\nu$ is consistent with  probing hotter chromospheric layers at higher heights, indicating the existence of a chromosphere and, with that, the presence of chromospheric heating and potentially higher activity (Paper I).  
In this paper, we fit the compiled mm-\Tbp\ for every star and explore the relationships between spectral structure and stellar physical properties, particularly focussing on the cool stars in the sample. 

In Sect.~\ref{sec:method},   the data and analysis methodology are described. 
The results are presented in Sect.~\ref{sec:results}, and a discussion in Sect.~\ref{sec:discussion} is followed by our conclusions  and outlook in  Sect~\ref{sec:conclusion}.


\section{Data and methodology} 
\label{sec:method}
The mm-\Tbp\ data used for this work are part of the sample of main-sequence stars presented in Paper~I. 
Here, we selected only those stars in the sample with more than one data point in the mm/sub-mm range (30 -- 1000\,GHz). The resultant sub-sample of nine stars covers a \teff\ range of $\sim$ 5000 -- 10,0000\,K (A -- K type).
Though the original sample compiled data in the 10 -- 1000\,GHz range, we chose to stick to the mm/sub-mm range for two reasons. 
{Firstly, for cool stars, the assumption of LTE emission is believed to be generally valid beyond 30\,GHz. 
For the Sun, this assumption is confirmed by detailed 3D atmospheric model calculations \citep[see][and references therein]{Sven16_ALMA_science}. 
For the other cool stars in our sample, the LTE nature of the mm emission has been demonstrated via data-driven modelling studies using scaled 1D solar models \citep{VAL1981} or solar-like models \citep{2020ApJS..246....5T}. \citet{2018MNRAS.481..217T} carried out this exercise for the $\alpha$\,Cen system (G and K dwarf) and \citet{White20_MESAS} did so for the F dwarfs.
However, below 30\,GHz, non-thermal emission from the corona can be relevant as shown for $\epsilon$\,Eri by 
\citet{suresh20_EpsEri_RadioSEDmodel}.} 
Secondly, the data below 30\,GHz are either very sparse or non-existent for the stars in our sample.

For the $\alpha$\,Cen binary, there exist multiple ALMA observations at certain frequencies gathered years apart and hence revealing the inherent time variability in the emission. 
Similarly $\epsilon$\,Eri observations at different frequencies were made years apart \citep{suresh20_EpsEri_RadioSEDmodel}. The solar data at 34\,GHz come from periods of minimum and maximum activity during solar cycle 23 \citep{2004NewAR..48.1319W}.
So, the mm-\Tbp\ of these stars represent emission averaged over the respective activity cycle timescales.

\section{Results}
\label{sec:results}
{Figure~\ref{fig1:fits} shows a collage of \Tbp\ for the cool stars in our sub-sample along with an A-type star for comparison. The years during which data were collected are given in the legends along with the corresponding radio telescopes. Multi-epoch observations at the same frequency are shown in different colours.}
The dotted line shows the photospheric emission model.
{For the Sun, quiet chromospheric spectra based on FAL A and C models \citep{2004A&A...419..747L} are shown, revealing the chromospheric origin of the observed \Tbp.
The observed \Tbp\ of every star was fitted by a power law (\Tbp\ $\propto \nu^{\alpha_{mm}}$). The shaded region in each subplot shows the family of curves consistent within the error range of the $\chi^2$ fit. 
The errors on \alfB\ could not be estimated for $\gamma$\,Vir\,A and B 
since there are only two \TB\ data points.  
While deriving $\alpha_{mm}$ for the $\alpha$\,Cen system and the Sun, we used mean \TB\ values at frequencies for which multi-epoch data existed. A systematic error was added to the mean \TB\ to account for emission variability.}
Despite the emission variability in the multi-epoch data, \alfB\ values are quite robust with relatively low-error ranges, letting us explore its physical dependencies.
\begin{figure}[!t]
\centering
\includegraphics[width=0.49\textwidth, height=0.28\textheight]{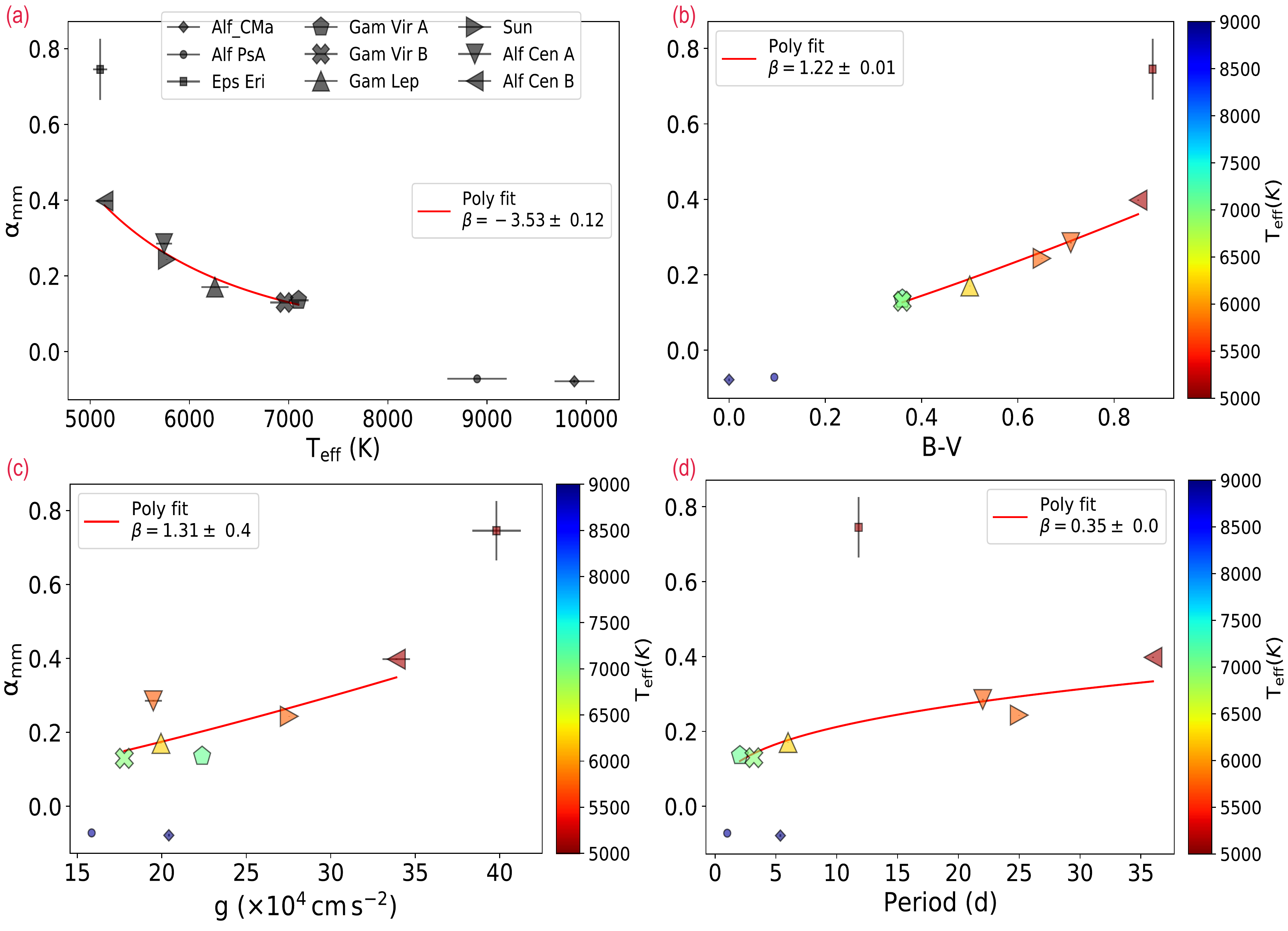}
\caption{\alfB\ versus physical parameters with respective power-law fits for the old group ($>$1\,Gyr). Smaller markers denote stars $<$1\,Gyr.}
\label{fig2:alf_Vs_pars}
\vspace{-0.5cm}
\end{figure}
\begin{figure}[!t]
\centering
\includegraphics[width=0.49\textwidth, height=0.15\textheight]{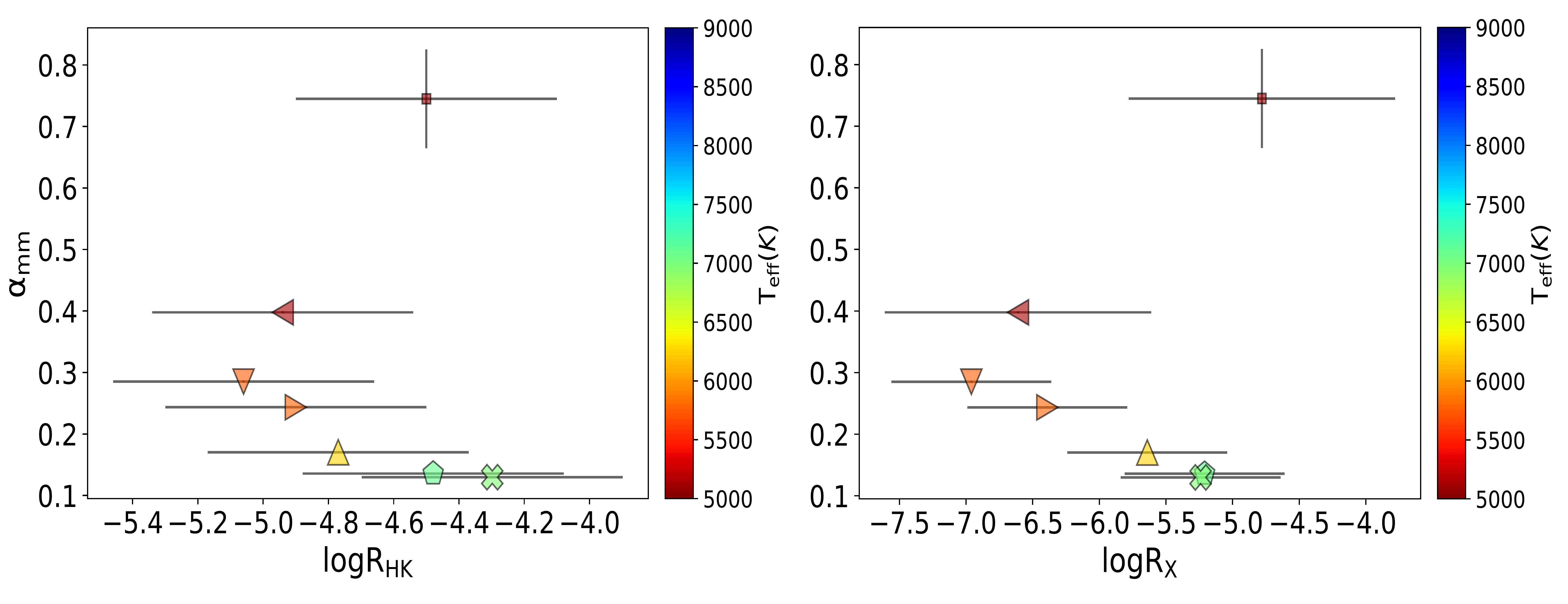}
\vspace*{-6mm}
\caption{Comparison of \alfB\ value ranges with those in \lRHK\ and \lRx\ for cool star sample. The markers used are the same as in Fig.~\ref{fig2:alf_Vs_pars}.}
\label{fig3:alfmm_Vs_AI}
\vspace{-0.5cm}
\end{figure}

Figure~\ref{fig2:alf_Vs_pars} explores \alfB\ versus various physical parameters.   
The sample consists of both an old and young group of stars with ages either above or below 1\,Gyr, respectively. 
The plot shows the old group in bigger markers, revealing the physical dependencies of \alfB, modelled by power-law functions (red curves) with the index, $\beta$.
There is only one cool star in the young group, $\epsilon$\,Eri. 
The A-type stars $\alpha$PsA and $\alpha$\,CMa show a negative \alfB, as they lack hotter upper atmospheres.
The \Tbp\ of A stars are consistent with the photospheric model, as seen in Fig.~\ref{fig1:fits} for $\alpha$\,CMa. We need more data for young stars to study their physical trends.

In addition, Figure~\ref{fig3:alfmm_Vs_AI} compares \alfB\ with common activity indicators. The error intervals on \lRx\ and \lRHK\ were estimated based on the results on their systematic variability for cool stars by \citet{2003A&A...404..637M} and \citet{Pace13_RHK_Drastic_variability}, respectively. 
The relatively low errors on \alfB\ help distinguish the cool stars much better than other indicators and they provide robust quantitative maps to physical parameters. 
\section{Discussion}
\label{sec:discussion}
Stars are believed to evolve from a high- to low-activity phase with age, due to continuous loss of energy and angular momentum, which is reflected in the rise of the rotation period with age \citep[e.g.][]{1972ApJ...171..565S, Stepien94_Ro_activity_rot_relations, Davenport16_keplerflares_age-flaring_rel, 2016Natur.529..181V}.
\citet{Barnes03_Rot_Vs_age_Vs_Activity} classified this age-related activity evolution by analysing the shift in the distribution of stars in a rotation period versus B-V plane for stars of different age ranges.
The two groups of distributions that emerged were named `I', the fast rotating high-activity group, and `C', the slow rotating low-activity group. 
It was found that stars tend to evolve from `C' to `I' as they age, with hotter stars evolving faster.
However, F -- K dwarfs older than 1\,Gyr were found to have migrated to the `I' branch. 
In our cool star sub-sample, we find that the old group shows robust physical trends with physical parameters, to which the lone young star, $\epsilon$\,Eri, does not agree. $\epsilon$\,Eri also shows a significantly higher \alfB.
Since \Tbp\ of cool stars is a proxy to their chromospheric thermal structure and since the spectral index, \alfB, quantifies the thermal gradients, the relatively high \alfB\ of $\epsilon$\,Eri 
hints at higher atmospheric heating and activity. 
\subsection{Physical dependencies of \alfB\ for old cool stars}
\label{sec:dic_sec1}
\begin{figure}[!t]
\centering
\includegraphics[width=0.5\textwidth, height=0.14\textheight]{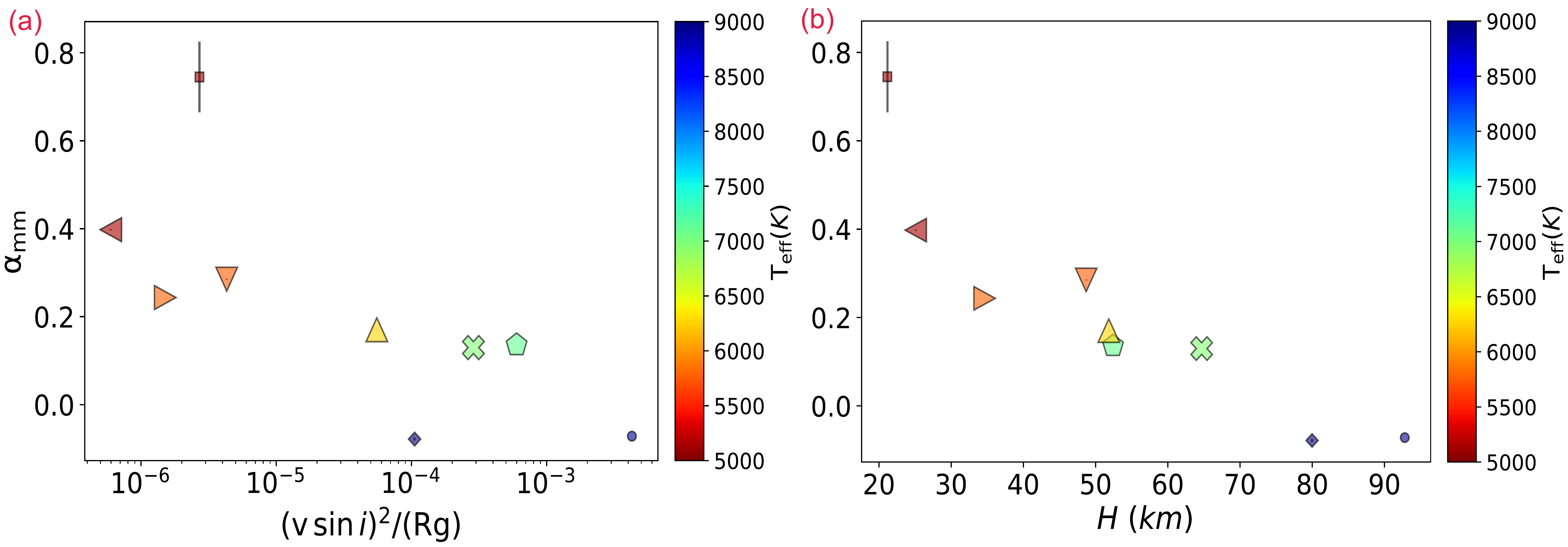}
\vspace*{-2mm}
\caption{\alfB\ versus (v\,sin$i$)$^2$/(Rg) and H. Markers are the same as in Fig.$\ref{fig2:alf_Vs_pars}$}
\label{fig4:alfB_vs_lngthscale}
\vspace{-0.5cm}
\end{figure}

Cooler stars (lower \teff) are expected to be more active due to their stronger magnetic fields. The \alfB\ estimates are higher for cooler stars, probably hinting at more efficient heating mechanisms which can sustain stronger chromospheric thermal gradients, as well as higher levels of steady heating and activity. Similarly, stars with higher B-V have higher \alfB.
We find that \alfB\ increases with period, as opposed to the well-known anti-correlation between activity and rotation period. 
Also, stars with higher surface gravity (g) show higher \alfB. 
{
Since activity is related to the observed rotation rates (v\,sin$i$; $i$ is the inclination angle with respect to the line of sight), g and \teff, analysing the individual parameter trends with \alfB\ may not provide a coherent picture, especially in view of the small sample size.

In Fig.~\ref{fig4:alfB_vs_lngthscale}(a), \alfB\ is compared to  (v\,sin$i$)$^2$/(Rg), the ratio of surface rotational kinetic energy relative to the gravitational potential. Stars with higher \alfB\ tend to show lower relative rotational kinetic energy. 
Figure~\ref{fig4:alfB_vs_lngthscale}(b) shows \alfB\ versus the scale height H = k\teff/($\mathrm{\mu m_p}$g), where k is the Boltzmann constant, $\mathrm{\mu}$ is the mean molecular weight (here $\mu =0.5$), and $\mathrm{m_p}$ is the proton mass. 
We find that \alfB\ increases with decreasing H.} 
Meanwhile, since $i$ values are unknown for the F dwarfs in the sample, the trends in the period and (v\,sin$i$)$^\mathrm{2}$/(Rg), will be affected by the projection effect (sin\,$i$). {Also we need multi-frequency data for the F dwarfs in the sample to make a robust \alfB\ estimate. A large multi-frequency sample of A- to M-type stars, with known $i$, spread across wide ranges in age, \teff, period, and g is needed to draw statistically significant conclusions. The A stars in such a sample will help explore the physics of emergence of hot chromospheres \citep{2002ApJ...579..800S} traced by \alfB\ becoming positive.} 

\subsection{Towards a robust mm-band activity indicator}
A robust activity indicator should not only provide a reliable quantitative scale, but also map the activity levels to physical parameters robustly, in a way that it facilitates clear segregation of stars in the planes of activity versus different physical parameters.
{ We considered two pairs} of stars with quite similar \teff\ and spectral types, and a good estimation of \alfB\ errors: $\alpha$\,Cen B and $\epsilon$\,Eri (Pair 1: K type, \teff\ $\sim$5000\,K), and $\alpha$\,Cen A and the Sun (Pair 2: G type, \teff\ $\sim$ 5800\,K). 
The stars in each pair differ among each other in physical parameters other than \teff.
Each pair of stars is well separated along the  \alfB\ axis, despite estimation errors.
In the case of Pair 1, $\epsilon$\,Eri belonging to the more active `C' branch has relatively high T$_B$ values for similar frequencies and higher \alfB\ than $\alpha$\,Cen\,B. 
Meanwhile, for Pair 2 comprised of old stars, in planes formed by their differentiating physical parameters (g, rotation period, B-V) with \alfB\ (Fig.~\ref{fig2:alf_Vs_pars}b -- d), the stars align well along the best-fit power-law trends.
This property of alignment along best-fit trends is even seen for the binary pair $\gamma$\,Vir\,A and B which have similar physical properties, except for g. $\gamma$\,Vir\,A with a slightly higher g has a slightly higher \alfB\ than its companion. 
A physical characterisation of cool stars to this accuracy is difficult with common indicators due to the large value intervals (Fig.~\ref{fig3:alfmm_Vs_AI}), and especially for stars within the narrow \teff\ range of 5000 - 7000\,K explored here.

As mentioned earlier, the error intervals on \alfB\ capture the effect of stellar variability within activity cycle periods for the Sun, $\epsilon$\,Eri, and the $\alpha$\,Cen binary.
It is found that the error interval is the highest for the most active young star, $\epsilon$\,Eri, possibly hinting at a higher variability in its activity level during the stellar activity cycle.
This agrees well with the current consensus that young `C' branch stars usually show higher variability in activity indicators than the older `I' branch during their activity cycles \citep[e.g.][]{Pace13_RHK_Drastic_variability}.
So, while the \alfB\ estimates provide a quantitative physical characterisation of the mean stellar activity, the error intervals reflect the variability across activity cycles.
We need more multi-epoch data for stars in different activity branches to explore mm-\Tbp\ and \alfB\ variability.

\section{Conclusions and outlook}
\label{sec:conclusion}
{It is difficult to derive robust quantitative correlations between stellar activity and physical parameters using the common activity indicators. 
This study explores the possibility of defining a robust activity indicator based on continuum radiation at  mm wavelengths  
using its unique tomographic potential  to trace the thermal emission from the various stellar chromospheric layers as a function of observing frequency.
}

The mm stellar sample compiled by \citet{2021A&A...655A.113M} was used for the study. 
The compiled \Tbp\ data come from observations done over years, providing chromospheric thermal profiles averaged over activity cycle timescales.
The cool stars showed a spectral steepening towards lower frequencies, as expected from an atmosphere that gets progressively hotter towards higher heights (explored by lower frequencies).
We derived the mm-\Tbp\ spectral index, \alfB, for the cool stars, which is expected to characterise the chromospheric thermal gradients and in turn the efficiency of atmospheric heating and activity.
Robust power-law functions could be derived for \alfB\ with different activity influencing physical parameters, namely \teff, g, and the rotation period. 
{In addition, \alfB\ shows an inverse correlation with the scale height and the ratio of surface rotational kinetic to potential energy; however, this finding should be interpreted with caution in view of the small sample size.}
Due to insufficient data, this analysis could only be done for the cool stars older than 1\,Gyr. 
However, the young cool star, $\epsilon$\,Eri, is well apart from the trends of the old group.

The estimation errors on \alfB\ are quite small in comparison to those in common activity indicators, which facilitated the robust activity quantification and mapping to physical parameters.
Even the stars with similar values for some physical parameters, such as the rotation period and \teff\,, among others, could be well separated in the \alfB\ axes.
{Also, with more data for A-type stars, the emergence of chromospheres can be explored by tracking the change of sign in \alfB\ as a function of physical parameters.}

To conclude, \alfB\ could be developed into a robust activity indicator giving  us vital insights into the links between stellar activity, atmospheric structure, and physical properties, especially for cool stars. To assert this, we need more observations and preferably a campaign to monitor nearby cool stars in the mm band at multiple epochs sufficient to sample their variability within respective activity cycles. 

\begin{acknowledgements}
This work is supported by the Research Council of Norway through the EMISSA project (project number 286853) and the Centres of Excellence scheme, project number 262622 (``Rosseland Centre for Solar Physics'').  
We acknowledge helpful discussions among the ISSI international team 387 ``A New View of the Solar-stellar Connection with ALMA'', which was funded by the International Space Science Institute (ISSI, Bern, Switzerland). 
This paper makes use of the following ALMA data: ADS/JAO.ALMA\#2013.1.00170.S, \#2016.1.00441.S, \#2013.1.00645.S.
ALMA is a partnership of ESO (representing its member states), NSF (USA) and NINS (Japan), together with NRC(Canada), MOST and ASIAA (Taiwan), and KASI (Republic of Korea), in co-operation with the Republic of Chile. The Joint ALMA Observatory is operated by ESO, AUI/NRAO and NAOJ. We thank the referee, Jeffrey L. Linsky, for his valuable comments and suggestions. 

\end{acknowledgements}

\bibliographystyle{aa}
\bibliography{EMISSA_allref}

\begin{thebibliography}{35}
\expandafter\ifx\csname natexlab\endcsname\relax\def\natexlab#1{#1}\fi

\bibitem[{Barnes(2003)}]{Barnes03_Rot_Vs_age_Vs_Activity}
Barnes, S.~A. 2003, \apj, 586, 464

\bibitem[{{Bashi} {et~al.}(2020)}]{Bashi20_occurrence_of_smallplanetsFGK}
{Bashi}, D. {et~al.} 2020, \aap, 643, A106

\bibitem[{Davenport(2016)}]{Davenport16_keplerflares_age-flaring_rel}
Davenport, J. R.~A. 2016, \apj, 829, 23

\bibitem[{{De Pascale} {et~al.}(2014)}]{HARPS2014_datarelease}
{De Pascale}, M. {et~al.} 2014, \aap, 570, A68

\bibitem[{{Donati} \& {Landstreet}(2009)}]{Donati09_Rev_Bfield}
{Donati}, J.~F. \& {Landstreet}, J.~D. 2009, \araa, 47, 333

\bibitem[{{Gaia Collaboration}(2021)}]{GaiaDR32021}
{Gaia Collaboration}. 2021, \aap, 649, A1

\bibitem[{{Grie{\ss}meier} {et~al.}(2005){Grie{\ss}meier}, {Stadelmann},
  {Motschmann}, {Belisheva}, {Lammer}, \& {Biernat}}]{2005AsBio...5..587G}
{Grie{\ss}meier}, J.~M., {Stadelmann}, A., {Motschmann}, U., {et~al.} 2005,
  Astrobiology, 5, 587

\bibitem[{{Hauschildt} \& {Baron}(1999)}]{1999JCoAM.109...41H}
{Hauschildt}, P.~H. \& {Baron}, E. 1999, \jcom, 109, 41

\bibitem[{{Lammer} {et~al.}(2007){Lammer}, {Lichtenegger}, {Kulikov},
  {Grie{\ss}meier}, {Terada}, {Erkaev}, {Biernat}, {Khodachenko}, {Ribas},
  {Penz}, \& {Selsis}}]{2007AsBio...7..185L}
{Lammer}, H., {Lichtenegger}, H. I.~M., {Kulikov}, Y.~N., {et~al.} 2007,
  Astrobiology, 7, 185

\bibitem[{Linsky(2017)}]{Linsky16_Stellar_chromRev}
Linsky, J.~L. 2017, Annual Review of Astronomy and Astrophysics, 55, 159

\bibitem[{{Loukitcheva} {et~al.}(2004){Loukitcheva}, {Solanki}, {Carlsson}, \&
  {Stein}}]{2004A&A...419..747L}
{Loukitcheva}, M., {Solanki}, S.~K., {Carlsson}, M., \& {Stein}, R.~F. 2004,
  \aap, 419, 747

\bibitem[{{Marsden} {et~al.}(2014){Marsden}, {Petit}, {Jeffers}, {Morin},
  {Fares}, {Reiners}, {do Nascimento}, {Auri{\`e}re}, {Bouvier}, {Carter},
  {Catala}, {Dintrans}, {Donati}, {Gastine}, {Jardine}, {Konstantinova-Antova},
  {Lanoux}, {Ligni{\`e}res}, {Morgenthaler}, {Ram{\`\i}rez-V{\`e}lez},
  {Th{\'e}ado}, {Van Grootel}, \& {BCool Collaboration}}]{2014MNRAS.444.3517M}
{Marsden}, S.~C., {Petit}, P., {Jeffers}, S.~V., {et~al.} 2014, \mnras, 444,
  3517

\bibitem[{{Micela} \& {Marino}(2003)}]{2003A&A...404..637M}
{Micela}, G. \& {Marino}, A. 2003, \aap, 404, 637

\bibitem[{{Mohan} {et~al.}(2021){Mohan}, {Wedemeyer}, {Pandit}, {Saberi}, \&
  {Hauschildt}}]{2021A&A...655A.113M}
{Mohan}, A., {Wedemeyer}, S., {Pandit}, S., {Saberi}, M., \& {Hauschildt},
  P.~H. 2021, \aap, 655, A113

\bibitem[{{Moschou} {et~al.}(2019){Moschou}, {Drake}, {Cohen},
  {Alvarado-G{\'o}mez}, {Garraffo}, \& {Fraschetti}}]{2019ApJ...877..105M}
{Moschou}, S.-P., {Drake}, J.~J., {Cohen}, O., {et~al.} 2019, \apj, 877, 105

\bibitem[{{Noyes} {et~al.}(1984){Noyes}, {Hartmann}, {Baliunas}, {Duncan}, \&
  {Vaughan}}]{Noyes84_RHK}
{Noyes}, R.~W., {Hartmann}, L.~W., {Baliunas}, S.~L., {Duncan}, D.~K., \&
  {Vaughan}, A.~H. 1984, \apj, 279, 763

\bibitem[{{Odert} {et~al.}(2020){Odert}, {Leitzinger}, {Guenther}, \&
  {Heinzel}}]{2020MNRAS.494.3766O}
{Odert}, P., {Leitzinger}, M., {Guenther}, E.~W., \& {Heinzel}, P. 2020,
  \mnras, 494, 3766

\bibitem[{{Pace}(2013)}]{Pace13_RHK_Drastic_variability}
{Pace}, G. 2013, \aap, 551, L8

\bibitem[{{Perryman} {et~al.}(1997){Perryman}, {Lindegren}, {Kovalevsky},
  {Hoeg}, {Bastian}, {Bernacca}, {Cr{\'e}z{\'e}}, {Donati}, {Grenon},
  {Grewing}, {van Leeuwen}, {van der Marel}, {Mignard}, {Murray}, {Le Poole},
  {Schrijver}, {Turon}, {Arenou}, {Froeschl{\'e}}, \&
  {Petersen}}]{1997A&A...323L..49P}
{Perryman}, M.~A.~C., {Lindegren}, L., {Kovalevsky}, J., {et~al.} 1997, \aap,
  323, L49

\bibitem[{{Simon} {et~al.}(2002){Simon}, {Ayres}, {Redfield}, \&
  {Linsky}}]{2002ApJ...579..800S}
{Simon}, T., {Ayres}, T.~R., {Redfield}, S., \& {Linsky}, J.~L. 2002, \apj,
  579, 800

\bibitem[{{Skumanich}(1972)}]{1972ApJ...171..565S}
{Skumanich}, A. 1972, \apj, 171, 565

\bibitem[{{Stepien}(1994{\natexlab{a}})}]{stepien94_Defn_Rx+Ro_Vs_activity_n_manyCorCurves}
{Stepien}, K. 1994{\natexlab{a}}, \aap, 292, 191

\bibitem[{{Stepien}(1994{\natexlab{b}})}]{Stepien94_Ro_activity_rot_relations}
{Stepien}, K. 1994{\natexlab{b}}, \aap, 292, 191

\bibitem[{{Suresh} {et~al.}(2020){Suresh}, {Chatterjee}, {Cordes}, {Bastian},
  \& {Hallinan}}]{suresh20_EpsEri_RadioSEDmodel}
{Suresh}, A., {Chatterjee}, S., {Cordes}, J.~M., {Bastian}, T.~S., \&
  {Hallinan}, G. 2020, \apj, 904, 138

\bibitem[{{Tapia-V{\'a}zquez} \& {De la Luz}(2020)}]{2020ApJS..246....5T}
{Tapia-V{\'a}zquez}, F. \& {De la Luz}, V. 2020, \apjs, 246, 5

\bibitem[{{Trigilio} {et~al.}(2018){Trigilio}, {Umana}, {Cavallaro},
  {Agliozzo}, {Leto}, {Buemi}, {Ingallinera}, {Bufano}, \&
  {Riggi}}]{2018MNRAS.481..217T}
{Trigilio}, C., {Umana}, G., {Cavallaro}, F., {et~al.} 2018, \mnras, 481, 217

\bibitem[{{van Saders} {et~al.}(2016){van Saders}, {Ceillier}, {Metcalfe},
  {Silva Aguirre}, {Pinsonneault}, {Garc{\'\i}a}, {Mathur}, \&
  {Davies}}]{2016Natur.529..181V}
{van Saders}, J.~L., {Ceillier}, T., {Metcalfe}, T.~S., {et~al.} 2016, \nat,
  529, 181

\bibitem[{{Vernazza} {et~al.}(1981){Vernazza}, {Avrett}, \& {Loeser}}]{VAL1981}
{Vernazza}, J.~E., {Avrett}, E.~H., \& {Loeser}, R. 1981, \apjs, 45, 635

\bibitem[{{Vidotto} {et~al.}(2014){Vidotto}, {Gregory}, {Jardine}, {Donati},
  {Petit}, {Morin}, {Folsom}, {Bouvier}, {Cameron}, {Hussain}, {Marsden},
  {Waite}, {Fares}, {Jeffers}, \& {do Nascimento}}]{Vidotto14_B_Vs_age_n_rot}
{Vidotto}, A.~A., {Gregory}, S.~G., {Jardine}, M., {et~al.} 2014, \mnras, 441,
  2361

\bibitem[{{Vidotto} {et~al.}(2013)}]{Vidotto13_highB_forclosebyplanets}
{Vidotto}, A.~A. {et~al.} 2013, \aap, 557, A67

\bibitem[{{Wedemeyer} {et~al.}(2016){Wedemeyer}, {Bastian}, {Braj{\v{s}}a},
  {Hudson}, {Fleishman}, {Loukitcheva}, {Fleck}, {Kontar}, {De Pontieu},
  {Yagoubov}, {Tiwari}, {Soler}, {Black}, {Antolin}, {Scullion}, {Gun{\'a}r},
  {Labrosse}, {Ludwig}, {Benz}, {White}, {Hauschildt}, {Doyle}, {Nakariakov},
  {Ayres}, {Heinzel}, {Karlicky}, {Van Doorsselaere}, {Gary}, {Alissandrakis},
  {Nindos}, {Solanki}, {Rouppe van der Voort}, {Shimojo}, {Kato},
  {Zaqarashvili}, {Perez}, {Selhorst}, \& {Barta}}]{Sven16_ALMA_science}
{Wedemeyer}, S., {Bastian}, T., {Braj{\v{s}}a}, R., {et~al.} 2016, \ssr, 200, 1

\bibitem[{{White} {et~al.}(2020){White}, {Tapia-V{\'a}zquez}, {Hughes},
  {Mo{\'o}r}, {Matthews}, {Wilner}, {Aufdenberg}, {Hughes}, {De la Luz}, \&
  {Boley}}]{White20_MESAS}
{White}, J.~A., {Tapia-V{\'a}zquez}, F., {Hughes}, A.~G., {et~al.} 2020, \apj,
  894, 76

\bibitem[{{White}(2004)}]{2004NewAR..48.1319W}
{White}, S.~M. 2004, \nar, 48, 1319

\bibitem[{{White} {et~al.}(2017){White}, {Iwai}, {Phillips}, {Hills}, {Hirota},
  {Yagoubov}, {Siringo}, {Shimojo}, {Bastian}, {Hales}, {Sawada}, {Asayama},
  {Sugimoto}, {Marson}, {Kawasaki}, {Muller}, {Nakazato}, {Sugimoto},
  {Braj{\v{s}}a}, {Skoki{\'c}}, {B{\'a}rta}, {Kim}, {Remijan}, {de Gregorio},
  {Corder}, {Hudson}, {Loukitcheva}, {Chen}, {De Pontieu}, {Fleishmann},
  {Gary}, {Kobelski}, {Wedemeyer}, \& {Yan}}]{2017SoPh..292...88W}
{White}, S.~M., {Iwai}, K., {Phillips}, N.~M., {et~al.} 2017, \solphys, 292, 88

\bibitem[{{Zendejas} {et~al.}(2010){Zendejas}, {Segura}, \&
  {Raga}}]{2010Icar..210..539Z}
{Zendejas}, J., {Segura}, A., \& {Raga}, A.~C. 2010, \icarus, 210, 539

\end{thebibliography}

\appendix

\end{document}